\documentclass[11pt, nofootinbib, showpacs]{revtex4-1}
\usepackage{amsmath}
\usepackage{graphicx}
\usepackage{wasysym}
\usepackage{amssymb}
\usepackage{epsf}
\usepackage{bm}
\usepackage{changes}

\def\apj{ApJ}
\def\apjl{ApJ}

\def\jcap{J. Cosmol. Astropart. Phys.}

\def\sovast{SvA}

\def\mnras{MNRAS}

\def\pasa{Publ. Astron. Soc. Australia}

\def\prd{Phys. Rev. D.}

\usepackage{graphicx}
\usepackage{graphicx}
\begin{document}
\title{Constraints on ultralight scalar dark matter from pulsar-timing}

\author{N.K. Porayko}
\email[]{porayko.nataliya@gmail.com}
\affiliation{Faculty of Physics, Moscow M.V. Lomonosov State University, Moscow, Russia}
\author{K.A. Postnov}
\email[]{pk@sai.msu.ru}
\affiliation{Faculty of Physics and Sternberg Astronomical Institute, Moscow M.V. Lomonosov State University, Moscow, Russia}

\date{\today}

\begin{abstract}
We perform a Bayesian analysis of pulsar-timing residuals from the NANOGrav 
pulsar-timing array to search for a specific form of stochastic 
narrow-band signal produced by oscillating gravitational potential (Gravitational Potential Background)
in the galactic halo. Such oscillations arise in models 
of warm dark matter composed of an ultralight massive 
scalar field ($m\sim 10^{-23}$~eV), recently considered by Khmelnitsky and Rubakov [J. Cosmol. Astropart. Phys. 2(2014)019].
In the monochromatic approximation, 
the stringent upper limit (95\% C.L.) on the variable gravitational potential 
amplitude is found to be  $\Psi_c<1.14 \times 10^{-15}$, 
corresponding to the characteristic strain 
$h_c=2\sqrt{3}\Psi_c < 4\times 10^{-15}$ at $f=1.75\times10^{-8}\mathrm{Hz}$.
In the narrow-band approximation, the upper limit of this background 
energy density is 
\(\Omega_{\mathrm{GPB}}<1.27 \times 10^{-9}\) at \(f=6.2\times 10^{-9}\mathrm{Hz}\). 
These limits are an order of magnitude higher than the expected signal amplitude. 
The applied analysis of the pulsar-timing residuals can be used to search for any 
narrow-band stochastic signals with different correlation properties.
As a by-product, parameters of the red noise present in four NANOGrav pulsars
(J1713+0747, J2145-0750, B1855+09, and J1744-1134)
have been evaluated.
\end{abstract}

\maketitle

\section{Introduction}
\label{Introduction}
Gravitational waves (GWs), predicted by general relativity, remain major,
directly nondetected fundamental spacetime features. The indirect evidence
for their existence was firmly obtained by measurements of the orbital decay in 
binary pulsars, which are in agreement with general relativity to better than half a percent
\cite{2014arXiv1403.7377W}. Recently, the trace of a primordial stochastic GW background, directly relating to the tensorial nature of GWs, 
was possibly found in the BICEP2 polarization measurements \cite{2014arXiv1403.3985B}, 
which strongly boosted interest in GW astronomy.     
Prompt development of GW detectors and projects, including ground-based and space interferometers, pulsar-timing and measurement of the anisotropy of cosmic microwave background, will likely result in the direct detection of GWs in the near future (see recent reviews \cite{lrr-2013-7, *lrr-2013-9}). 

Pulsars, which are rapidly rotating neutron stars with highly stable spin frequency, are recognized to be sensitive GW detectors 
\cite{Sazsin, *Det}. Especially suitable for GW detection are millisecond pulsars --- old neutron stars
spun up to millisecond periods due to accretion in binary systems \cite{lrr-2008-8}.
A pulsar-timing GW detector is represented by two ``free'' masses: 
Earth and a pulsar. Propagation of a GW induces a weak imprint (through the Doppler shift) 
in the times of arrival (TOA) of pulses emitted from the pulsar \cite{Estabrook}. Potentially, these imprints could be measured by application of statistical methods to the so-called timing residuals (i.e. the difference between the observed and model-predicted TOA). However, the TOA are also influenced by uncertainties in the sky location of the pulsar, model characteristics of the pulsar companion (in the case of binary pulsars \cite{1999MNRAS.305..563K}), the radio beam propagation effects through the ionized interstellar medium, etc. 
The pulsar-timing analysis takes into account these model parameters and thus can be used for a more accurate determination of the physical model of the pulsar itself. 

The pulsar-timing procedure is sensitive to the GWs in the frequency range limited by the Nyquist frequency (as determined by the duty cycle of the measurements, about two weeks) 
and by the whole time span of the observations (usually several years), i.e. 
\(f_{\text{GW}}\in[10^{-9}\mathrm{Hz};10^{-7}\mathrm{Hz}]\). In this frequency range, potential astrophysical GW sources include supermassive black hole binaries (SMBHBs) \cite{Jenet}, which can be located in the centers of galaxies, and a stochastic gravitational wave background (GWB) produced by the whole population of SMBHBs \cite{Sesana} or, likely, by several bright sources above a weak GWB \cite{Bab, *BabakSes}. 
The GW detection procedure is also determined by properties of the sought signal. 
In the simplest case, in TOA from one pulsar a monochromatic plane GW with 
amplitude $h_c$ and frequency $f$ produces an oscillatory timing residuals, which are also
determined by the pulsar distance $D$ and the 
relative position of the GW source and the pulsar (via the angle between the direction to the source and the pulsar).  

Cross correlation of residuals from
different pulsars 
can be used to search for 
stochastic GW signals as well \cite{Hellings}.  
This concept forms the basis for the construction of pulsar-timing Arrays (PTAs), which nowadays are brought about in EPTA \cite{Janssen}, PPTA \cite{Manchest}, NANOGrav \cite{Nanoam}, and joints in IPTA \cite{IntPuls} (see review of the PTA techniques in Ref. \cite{2013CQGra..30v4001L}). In the last years, the PTA technique resulted in astrophysically interesting upper limits 
on GW signals of different kinds in the frequency range $10^{-9}-10^{-7}$~Hz (e.g., Ref. \cite{2010MNRAS.407..669Y, *2014arXiv1404.1267A}).

In addition to the ``traditional'' GW sources and stochastic backgrounds that can be probed in the PTA frequency range, 
there can be more exotic signals, including, for example, GWs from oscillating string loops \cite{Damour}, GW signals with memory \cite{Postnov, *Levinmem}, and GWs from massive gravitons \cite{2005PhRvL..94r1102D, Baskaran, Pshirkov, LeeJen}. Recently, Khmelnitsky and Rubakov \cite{Khmelitsky} considered a model of an 
ultralight scalar field with mass $m\sim 10^{-23}$~eV 
that can be a viable warm dark matter candidate. 
Different aspects of ultralight scalar fields as warm 
dark matter have been discussed in the literature;
see e.g. Ref. \cite{1985MNRAS.215..575K, *2000PhRvL..85.1158H, *2002PhRvD..65h3514A, *2013arXiv1302.0903S, *PhysRevD.62.103517} and references therein. In the galactic halo, due to a huge occupation number, such a field 
has a coherent part that behaves like  a classical wave with amplitude $\sim \sqrt{\rho_{DM}}/m $ and
coherence time $\sim 2\pi/(m v^2)$, where $\rho_{DM}\approx 0.3$~GeV/cm$^3$ is the local 
dark matter density and $v\sim 300$~km/s is the virial halo velocity. As shown in Ref. \cite{Khmelitsky}, through purely gravitational coupling such a field would produce oscillations of the gravitational potential in the galactic halo at 
the frequency twice the field mass ($\sim 10^{-8}$~Hz for $m\sim 10^{-23}$~eV), falling within the PTA frequency range.
 Similar to GWs, such oscillations can be sought after in the pulsar-timing as monochromatic oscillating residuals with an amplitude
corresponding to the characteristic GW strain $h_c\sim 10^{-15}$. Through a dilatonic coupling 
with the standard model particles, these oscillations can also be probed by atomic clock
experiments \cite{2014arXiv1405.2925A}. 

A distinctive feature of the pulsar-timing residuals due to 
oscillating gravitational potentials produced by a variable scalar field 
is that the amplitude of the TOA residuals should be independent of the pulsar location in the sky. Such a signal is also not a collection of monochromatic GWs with different amplitudes and phases from distant sources. Therefore, it is interesting to put constraints on the amplitude of 
this specific signal from the available PTA data, which is the main goal of the present paper.   
To this aim, we used publicly available pulsar-timing data from the NANOGrav Project, which is described in detail in Ref. \cite{Demorest}. 

The plan of the paper is as follows. In Sec.\ref{signatures} we introduce the form of the monochromatic signal and correlation matrix for the narrow-band stochastic signal formed by a variable gravitational potential background (GPB). In Sec. \ref{method} we perform the method of data processing based on the likelihood function in the Bayesian approach. In Sec. \ref{data} we describe the data that have been used. In Sec. \ref{results} we summarize our results.


\section{Signatures of a massive scalar field in pulsar-timing}
\label{signatures}

A recent paper \cite{Khmelitsky} considered signatures of a massive scalar field,
which can be a viable model for (warm) dark matter, in the pulsar-timing observations. 
The scalar field particles with 
mass $m\sim 10^{-23}$~eV moving with the galactic virial velocity 
$v=10^{-3}$ have a de Broglie wavelength of around 1 kpc, which 
allows one to describe the galactic halo dark matter in terms of 
an essentially classical field. 
The field oscillates with frequency $\approx m$ and can be represented as
a collection of almost monochromatic ($\Delta \omega/\omega\sim v^2\sim 10^{-6}$) 
plane waves, producing the  
oscillating pressure, and hence, through purely gravitational coupling, the variable 
gravitational potentials $h_{00}=2\Phi$ and $h_{ij}=-2\Psi\delta_{ij}$
(in the Newtonian conformal gauge)\footnote{Incidentally, we 
note that in Ref. \cite{Khmelitsky}, the scalar potential $\Psi$ [their Eq. (2.8)]
is initially taken with the minus sign $h_{ij}=+2\Psi \delta_{ij}$, opposite to the standard choice \cite{2011iteu.book.....G}.
Therefore, the sum of the scalar potentials $\Phi+\Psi$ arises in the gradient term in their Eq. (3.2), and not the difference, as in the standard literature. This, however, has no effect for
the pulsar-timing of interest here.} at frequency $\omega=2\pi f=2m$.
The propagation of an electromagnetic signal from a pulsar 
through the time-dependent spacetime
will leave an imprint in the pulsar-timing, much like a gravitational wave.
From the physical point of view, this is the classical Sachs-Wolfe effect 
\cite{1967ApJ...147...73S, *2007GReGr..39.1929S}. 
A derivation for the propagating electromagnetic signal in the special case of time-dependent
scalar potentials can be found in the textbook \cite{2011iteu.book.....G} [see Appendix A, where 
we sketch the derivation of Eq. (3.9) in Ref. \cite{Khmelitsky}].
The plausible frequency interval of the potential variations in the model considered
(\(10^{-9}-10^{-7}\mathrm{Hz}\))
can be probed by the current pulsar-timing array observations. 

Although both scalar potentials $\Phi$ and $\Psi$ generally contribute to the redshift 
of electromagnetic signal propagating from the pulsar to the observer, only 
the variable part of the potential $\Psi_c\cos(\omega t+\alpha)$ ($\alpha$ is the field phase)
can be probed by pulsar-timing. It is this part that nontrivially depends on the 
local dark matter density and the field mass, $\Psi_c\sim \rho_{\text{DM}}/m^2$ (see also the discussion below in Sec. \ref{sec:discussion}).

The form of the resulting signal in the pulsar-timing residuals reads \cite{Khmelitsky} (see also
Appendix A) \footnote{In this expression, the term in the signal redshift containing the integral of  
spatial gradients of the potentials along the ray is ignored; this term is suppressed by factor $v\sim 10^{-3}$ relative to the value of the potentials, but can be easily taken into account in the
PTA data analysis, its contribution being atttributed to the field phase uncertainty.}:
\begin{equation}
\label{e:R(t)}
R(t)=\frac{\Psi_c}{2 \pi f}\left\{(\sin(2 \pi f t+2 \alpha (\textbf{x}_e))-\sin(2 \pi f (t-D/c)+2 \alpha (\textbf{x}_p))\right\},
\end{equation}
where \(f\) is the frequency, $D$ is the distance to the pulsar, $c$ is the speed of light, \(\alpha(\textbf{x}_e)\) and \(\alpha(\textbf{x}_p)\) are the field phases on Earth and at the pulsar, respectively, and
$\Psi_c$ is the variable potential amplitude to be constrained from the PTA timing analysis. 

The structure of the timing residuals produced by the variable gravitational potential is reminiscent of that from a plane gravitational wave with amplitude $h_c\sim \Psi_c$ but, unlike the GW residuals, is independent of the angle
between the GW source and the pulsar. Below, we will refer to the first and second terms in 
Eq. (\ref{e:R(t)}) as the ``Earth-term'' and ``pulsar-term'', respectively.  

\subsection{Monochromatic approximation}
\label{s:monochrom}

The expected signal is concentrated within 
a very narrow frequency band $\delta f/f\sim v^2\sim 10^{-6}$, much 
smaller than the current PTA frequency resolution $\Delta f/f \sim 10^{-4}$, and therefore 
can be treated as monochromatic. Let us examine this case first, 
i.e., neglect the signal widening due to the final mass of the scalar field particles (see the next subsection). 
In this approximation, the signal to be searched for in the TOA 
residuals is given by Eq. (\ref{e:R(t)}). 

In the PTA technique, given large uncertainties in the pulsar distance estimates,  
it is common to operate only with Earth terms correlated between different pulsars. 
For example, the justification for dropping the pulsar term in the case of GWs from supermassive black hole binaries is that the pulsar terms add up at different frequencies and phases \cite{
2012ApJ...756..175E, Bab}. Here, 
we will analyze both cases (including and dropping the pulsar term);
as for the GPB produced by the scalar field, the Earth and pulsar terms arise in one frequency bin but still with a phase difference [see also the footnote 1]. Thus, the required signal forms $s(t)$
can be written as follows:
\begin{equation}
s(t)=R(t)=\begin{cases}
\cfrac{\Psi_c}{2\pi f}\mathrm{sin}(2\pi f t + 2 \alpha (\textbf{x}_e)),\ \text{E. term only}\\
\cfrac{2\Psi_c}{2\pi f}\mathrm{sin}(\alpha (\textbf{x}_e)+\cfrac{\pi f D}{c}-\alpha (\textbf{x}_p))\mathrm{cos}(2\pi f t+\alpha (\textbf{x}_e)+\alpha (\textbf{x}_p)-\cfrac{\pi f D}{c}),\  \text{E. and P. terms}
\end{cases}
\label{signal}
\end{equation}
Below, we will denote the effective phase angle due to the pulsar 
\(\theta\equiv \alpha(\textbf{x}_p)-\pi f D/c\), which is individual for each pulsar and is assumed to be uniformly distributed within the interval \([0, 2\pi]\). 
A distinguishing feature of such a monochromatic signal is 
the same amplitude for each pulsar in the array with no connection between 
their angular positions in the sky. 


\subsection{Narrow-band approximation}
\label{s:nB}

Now, consider the general case of a stochastic narrow-band signal, which is different from 
the monochromatic case from the point of view of data processing. 
This approach may be useful in searching for possible narrow-band stochastic signals 
in PTA data. In addition, in the frame of this approach, it is 
straightforward to relate the amplitude of the oscillating gravitational
potential $\Psi_c$ considered in the present paper 
to the widely used dimensionless power of a stochastic background
in the logarithmic frequency interval $\Omega_{\mathrm{GPB}}$.  
We will see that the narrow-band approach gives the same constraints on the 
signal considered as
the monochromatic treatment discussed above, as it should be.

In this approximation, the signal is treated as a narrow-band
stationary stochastic background with power contained within the frequency band $\delta f$. The properties of this background can be 
characterized in a way similar to a stochastic GWB, however,
some differences do arise due to different geometrical structures of
GWs and a variable gravitational potential signal. To see this difference,
it is instructive to start with reminding readers of the standard description of 
a stochastic GWB \cite{2001PhyU...44R...1G, Baskaran}. 
  
The properties of a stationary statistically homogeneous and isotropic 
gravitational wave field can be fully described by the metric power spectrum
$P_h(k)$ per logarithmic interval of the wave number \(k=2 \pi f/c\):
\begin{equation}
\langle h_s(k^i)h_{s'}^*(k^{'i})\rangle = \delta_{ss'}\delta^3(k^i-k^{'i})
\frac{P_h(k)}{16\pi k^3},
\label{e:h(t)}
\end{equation}
where the angular brackets denote ensemble averaging over all possible 
realizations, the mode functions $h_s(k^i)$ correspond to plane monochromatic waves, and
$s=1,2$ correspond to two linearly independent modes of polarization.

The dimensionless strain amplitude of the GW field can be defined as
\begin{equation}
h_c(k)^2=P_h(k)\,,
\end{equation}
and the rms amplitude of the GW field is   
\begin{equation}
\langle h^2 \rangle= \int P_h(k) d\log k \,.
\end{equation}
The characteristic strain $h_c$ fully characterizes the power 
of the signal. In the case of a
narrow-band signal concentrated within some theoretically prescribed 
interval $\delta k$, 
one may equivalently introduce the spectral amplitude $P_0$
\begin{equation}
\label{e:delta}
P_h(k')=\begin{cases}P_0, k<k'<k+\delta k\\
0, \hbox{in other cases}.
\end{cases}
\end{equation}
It can be related to the characteristic strain as 
\begin{equation}
\label{e:P0}
h_c^2= \langle h^2\rangle =P_0\delta f/f\,.
\end{equation}
However, if the frequency interval determined by 
the detector resolution $\Delta f\gg \delta f$, only $h_c^2$ can be probed.

It is also customary to relate the characteristic 
strain amplitude \(h_c(k)\) to the energy density of a stochastic background
per logarithmic frequency interval 
\begin{equation}
\rho_{\mathrm{GWB}}=(16\pi G)^{-1} 4\pi^{2}f^2 h_c^2,
\end{equation}
or, in dimensionless units, 
\begin{equation}
\Omega_{\mathrm{GWB}}=\frac{\rho_{\mathrm{GWB}}}{\rho_{cr}}=\frac{2\pi^2}{3H_0^2}f^2h_c^2,
\end{equation}
where the current critical density is 
\(\rho_{cr}=3 H_0^2/(8 \pi G)\) and $H_0$ is the present-day Hubble constant.

For the PTA data analysis, we will also need the spectrum $S(f)$ of the TOA residuals produced by 
the sought stochastic signal:
\begin{equation}
\langle R^2 \rangle=\int \frac{dk}{k}P(k)\tilde R^2(k) = \int_0^\infty S(f) df\,.
\end{equation}
[Here $\tilde R(k)$ is the transfer function between the GW field and the timing residuals \cite{Baskaran}.]
For example, in the case of an isotropic GWB, the transfer function is $\tilde R^2_{\mathrm{GWB}}(k)=1/(3k^2c^2)$, and  for 
the one-sided spectral density of the residuals, we obtain the well-known result
\begin{equation}
S_{\mathrm{GWB}}(f)=\frac{h_c^2}{12\pi^2 f^3}.
\label{SpectrGWB}
\end{equation}
When deriving this formula,
the averaging over the GW tensorial structure and polarization properties 
has been made. Repeating the derivation of the transfer 
function $\tilde R^2(k)$ as in Ref. \cite{Baskaran} for the sought signal from 
oscillating scalar gravitational potential $\Psi_c$ [Eq. (\ref{e:R(t)})], we arrive at 
\begin{equation}
S_{\mathrm{GPB}}(f)=\frac{\Psi_c^2}{\pi^2 f^3},
\label{Spectr}
\end{equation}
which is 12 times as high as Eq. (\ref{SpectrGWB}). 
Incidentally, this independently checks the relation between the equivalent GW characteristic 
strain $h_c$ and the amplitude of the varying potential $\Psi_c$ calculated in 
\cite{Khmelitsky} [see their Eq. (3.9)]: $h_c=2\sqrt{3}\Psi_c$. Therefore, 
in the narrow-band approximation, 
the amplitude $\Psi_c$ can be related to the parameter $\Omega_{\mathrm{GPB}}$ as follows:
\begin{equation}
\label{e:OmegaGPB}
\Omega_{\mathrm{GPB}}=\frac{8\pi^2}{H_0^2}f^2\Psi_c^2\,.
\end{equation}

The  PTA data analysis requires the knowledge of the covariance function  $C$ of the sought signal.
For a stochastic background, the variance covariance function $C$ 
is related to the signal spectral density \(S(f)\) via the Wiener-Khinchin theorem:
\begin{equation}
C(\tau)=\int^{\infty}_{0}S(f)\cos(\tau f)df\,.
\label{Cint}
\end{equation}
Using the equation for the one-sided spectral density [Eq. (\ref{Spectr})]
and performing the integration (see Appendix C), 
we obtain the following expression for \(C_{\mathrm{GPB}}\):
\begin{equation}
\label{e:CGPB}
C_{\mathrm{GPB}}(\tau_{ij})=\zeta_{\alpha\beta}\frac{\Psi_c^2 \delta f}{\pi^2 f^3} \mathrm{cos}(f\tau_{ij}),
\end{equation}
where \(\tau_{ij}=2\pi|t_i-t_j|\), \(i\) and \(j\) are indexes of TOA, and \(f\) is the central frequency of the GPB under study. Here, \(\zeta_{\alpha\beta}\) is the correlation term between pulsars (\(\alpha\), \(\beta\)). As discussed above, GPB oscillations will induce a sinusoidal signal in 
the TOA of each pulsar with the correlation which takes the simple form
(in contrast, for example, to the case of the GWB from merging SMBHBs):
\begin{equation}
\zeta_{\alpha\beta}=1/2(1+\delta_{\alpha\beta})\,.
\end{equation}
Here, the first and second terms arise due to the correlations between 
the pulsar term and the Earth term in Eq. (\ref{signal}),
respectively.

\section{Method of data analysis}
\label{method}

Because of the pulsar-timing data being not evenly sampled in time and the 
data containing a time-correlated red noise, we have applied 
a Bayesian technique developed in Ref. \cite{Levin}. Here, we will briefly remind readers of the main points.

Generally, pulsar-timing TOA \(\textbf{t}^{arr}\) can be represented by two components: deterministic and stochastic:
\begin{equation}
\textbf{t}^{arr}=\textbf{t}^{det}(\bm{\beta})+\bm{\delta} \textbf{t}.
\end{equation}
The deterministic part is characterized by the pulsar model parameters \(\bm{\beta}\). If the initial guess \(\bm{\beta_0}\) is good, the linear relation between the timing residuals and the uncertainty \(\bm{\zeta}=\bm{\beta}-\bm{\beta_0}\) is used.

In our case the random process part \(\bm{\delta} \textbf{t}\) is assumed to include three components: white instrumental noise with a diagonal covariance matrix \(C_{\mathrm{WN}}\), a red intrinsic noise characterized by matrix 
\(C_{\mathrm{RN}}\), which could be, for example, due to the irregular exchange of momentum between the superfluid component and the crust of the neutron star, and the stochastic 
background \(C_{\mathrm{GPB}}\) under study (in the narrow-band approximation). 
Therefore, the covariance matrix of the random process for the TOA of pulsars in the array 
 \(\bm{\delta t}\) includes three components: \(C=C_{\mathrm{WN}}+C_{\mathrm{RN}}+C_{\mathrm{GPB}}\)
which can be expressed analytically or semianalytically \cite{LevinH}:
\begin{equation}
C_{\mathrm{WN}}=\sigma^2_{\alpha, i} \delta_{\alpha \beta} \delta_{i j},
\end{equation}
\begin{equation}
\begin{split}
C_{\mathrm{RN}}=\delta_{\alpha\beta}A^2_{\mathrm{RN},\alpha}\left(\frac{1}{2\sqrt{3}\pi {\text{yr}}^{-1}}\right)^2 \left(\frac{\mathrm{yr}^{-1}}{f_L}\right)^{\gamma^{\alpha}_{\mathrm{RN}}-1}\\
[\Gamma(1-\gamma^{\alpha}_{\mathrm{RN}}) \sin \frac{\pi \gamma^{\alpha}_{\mathrm{RN}}}{2}(f_L \tau_{ij})^{\gamma_{\mathrm{RN}}^{\alpha}-1}-
\\
-\sum_{n=0}^{\infty} (-1)^n \frac{(f_L \tau_{ij})^{2n}}{(2n)!(2n+1-\gamma_{\mathrm{RN}}^{\alpha})}],
\end{split}
\end{equation}
\begin{equation}
C_{\mathrm{GPB}}=\zeta_{\alpha\beta}\frac{\Psi_c^2 \delta f}{\pi^2 f^3} \cos(f\tau_{ij}),
\label{C}
\end{equation}
where \(\Gamma\) is the gamma function.
Here, \(A_{\mathrm{RN}}\) and \(\gamma_{\mathrm{RN}}\) are the effective strain amplitude and the 
power-law index of the one-sided power spectral density of the red noise, respectively, and
\(\sigma_{\alpha,i}\) is the $i$th TOA observation error in the data of pulsar \(\alpha\), where \(i\in [1, n_{\alpha}]\) and \(\alpha\in [1, N]\) (\(N\) is the number of pulsars in the array, and
\(n_{\alpha}\) is the number of observations for pulsar \(\alpha\)).
The red-noise low-frequency cutoff \(f_L\) defines the lower limit in the integral (\(\ref{Cint}\)), providing the convergence for the red-noise indexes \(1<\gamma_{\mathrm{RN}}<7\).

In the time domain, we use a likelihood function in order to estimate the parameters of our model. According to the Bayesian approach, the likelihood function, in the Gaussian approximation,  after marginalizing over the unwanted pulsar-timing parameters \cite{Levin}, takes the following form:
\begin{equation}
\begin{split}
P(\boldsymbol{\delta} \textbf{t}|\bm{\phi})=\frac{1}{\sqrt{(2\pi)^{(n-m)}\mathrm{det} (G^{\mathrm{T}} C G)}}\\
 \mathrm{exp}(-\frac{1}{2}\bm{\delta} \textbf{t}^{\mathrm{T}} G (G^{\mathrm{T}} C G)^{-1} G^{\mathrm{T}} \bm{\delta} \textbf{t}).
\end{split}
\end{equation}
Here, \(n\) is the dimension of \(\bm{\delta} \textbf{t}\), $m$ is a whole number of the unwanted parameters, \(\bm{\phi}\) is the noise parameter vector, and \(G\) refers to the product of the so-called ``design matrix'' that can be obtained using the design matrix plugin of the {\small TEMPO2} software \cite{LevinH, Hobbs}. 

In searching for the deterministic signals (\(\ref{signal}\)), we have used the logarithmic likelihood ratio function (the ratio of likelihoods in the case where 
the signal is present to the case where the signal is absent):
\begin{equation}
\mathrm{log} \Lambda = \bm{\delta} \textbf{t}^{\mathrm{T}} G(G^{\mathrm{T}}CG)^{-1} G^{\mathrm{T}} \textbf{s}-1/2 \textbf{s}^{\mathrm{T}} G(G^{\mathrm{T}}CG)^{-1} G^{\mathrm{T}}\textbf{s},
\end{equation}
which depends on two parameters: the amplitude \(\Psi_c\) and the Earth phase \(\alpha(\textbf{x}_{e})\) of the scalar field when using the Earth term only, and on \(N+2\) parameters: the amplitude \(\Psi_c\), the Earth phase \(\alpha(\textbf{x}_{e})\) and  the phase \(\theta_\beta=\alpha(\textbf{x}_{p}^{\beta})-\pi f D_{\beta}/c\), if both the Earth and pulsar terms are included. A uniform distribution for \(\alpha \in [0;2 \pi]\) and \(\theta_\beta \in [0;2 \pi]\) is assumed.
\begin{figure}[htp]
\centering
\includegraphics[scale=1.00]{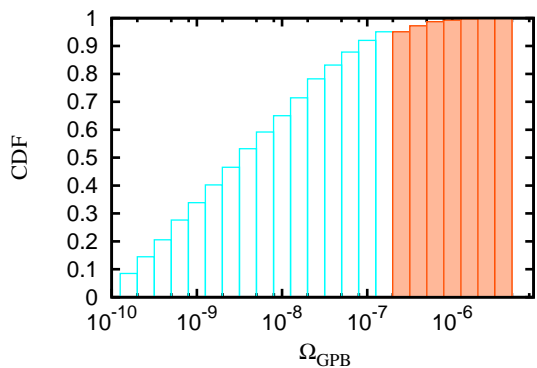}
\caption{(Color online). 
The typical form of the cumulative distribution of \(\Omega_{\mathrm{GPB}}\) 
in one frequency bin numerically reconstructed by the MCMC method. The blue color (light gray) shows the range containing 95\% of all Markov chain points. 
The quantile 0.95 of the \(\Omega_{\mathrm{GPB}}\) cumulative distribution function
gives the 95\% confidence level $\Omega_{\mathrm{GPB}}<\Omega_{0.95}$.}
\label{distribution}
\end{figure}

To obtain an upper limit on \(\Psi_c\) (\(\Omega_{\mathrm{GPB}}\)) as a function of the 
central frequency $f$, we split the entire interesting frequency range into small bins per logarithmic scale [\(\delta f/f\simeq 0.03 \ll 1/(5\,\mathrm{yrs})\)].

By assuming a uniform trial distribution for \(f\), \(\Psi_c\) and \(\Omega_{\mathrm{GPB}}\) 
and a normal trial distribution for \(\gamma_{\mathrm{RN}}\) and \(A_{\mathrm{RN}}\), we construct a long enough chain for each frequency bin using Markov chain Monte Carlo (MCMC) 
simulations Ref. \cite{Newman}. Taking into account the posterior distribution of \(\Psi_c\) (\(\Omega_{\mathrm{GPB}}\)), which is found to be close to a uniform distribution, we can estimate an upper limit as quantile 0.95 of the obtained \(\Psi_c\) (\(\Omega_{\mathrm{GPB}}\)) cumulative 
distribution (see Fig. \(\ref{distribution}\)). In other words, we estimate the posterior distribution of the amplitude with MCMC method and assume that 
the amplitude of the probable signal (even if the signal is present) with 95 \% probability lies 
within the 2-\(\sigma\) contour \cite{2011MNRAS.414.3117V}.
Results of this analysis are presented below in Sec. \ref{results}.

\section{Data description}
\label{data}
By applying the method described above, we have processed the real data from the NANOGrav Project. The observations, which are described in detail in Ref. \cite{DemorestNanoG} and are publicly available in \footnote{\url{http://www.cv.nrao.edu/~pdemores/nanograv_data/}}, 
were conducted using two radio telescopes, the Arecibo Observatory and the NRAO Green Bank Telescope. Each pulsar was observed nearly 30--60 days during a 5-yr period from 2005 to 2010. As the pulsar-timing array technique is not sensitive to GWs with one-day periods, we have conducted the procedure, depicted in Ref. \cite{Lommen}, to obtain the ``daily averaged'' TOA, in order to diminish the signal-to-noise ratio in each observation. The best results were obtained for PSRs J1713+0747 and J1909-3744 with a weighted rms \(\sim\)20-30ns. The search for a non-white-noise component in the NANOGrav data was performed in Ref. \cite{DemorestNanoG}.  For our purposes, we have chosen the observations of 12 pulsars in one wide frequency band for each pulsar: B1855+09(1400 MHz), J0030+0451(400 MHz), J0613-0200(800 MHz), J1012+5307(800 MHz), J1455-3330(800 MHz), J1600-3053(1400 MHz), J1640+2224(400 MHz), J1713+0747(1400 MHz), J1744-1134(800 MHz), J1909-3744(800 MHz), J1918-0642(800 MHz) and J2145-0750(800 MHz). Four of them (J1713+0747, J2145-0750, B1855+09 and J1744-1134) show a weak red-noise component that should be taken into account (see Appendix B). In searches for a monochromatic signal we have used only eight white-noise pulsars from the list, since the contamination of the sensitivity occurs due to unmodeled noise sources. In the narrow-band analysis, data for all 12 pulsars from the list were included.
The results of the analysis in the monochromatic and the narrow-band approximations 
were compared using only the eight white-noise pulsars. The postfit residuals were obtained with the {\small TEMPO2} software Ref. \cite{Hobbs}. The additive and multiplicative factors (EFAC, EQUAD) were not used in the data preprocessing \cite{LevinHaast}, so these parameters were not added to the 
``free parameter'' template in our model.

\section{Results}
\label{results}

In this section we present the results of searches for the GPB produced by 
massive scalar field oscillations in 
the NANOGrav pulsar-timing data.
Working in the time domain we have applied the Bayesian approach developed in Ref. \cite{Levin}. In order to estimate the red-noise parameters of pulsars, we have used MCMC method to find the distribution of the red-noise parameters, which were found to be non-Gaussian. In the data analysis, we 
examined three possible signal types: 
a monochromatic deterministic GPB with the Earth term only, a monochromatic GPB 
including both the Earth and pulsar terms, and a narrow-band stochastic GPB. In all cases, we 
obtained an upper limit on the signal amplitude \(\Psi_c\) (or \(\Omega_{\mathrm{GPB}}\)) as a function of frequency $f$. The best sensitivity is reached in the case of the monochromatic signal using both the Earth and pulsar terms.

\begin{figure}[htp]
\centering
\includegraphics[scale=1.20]{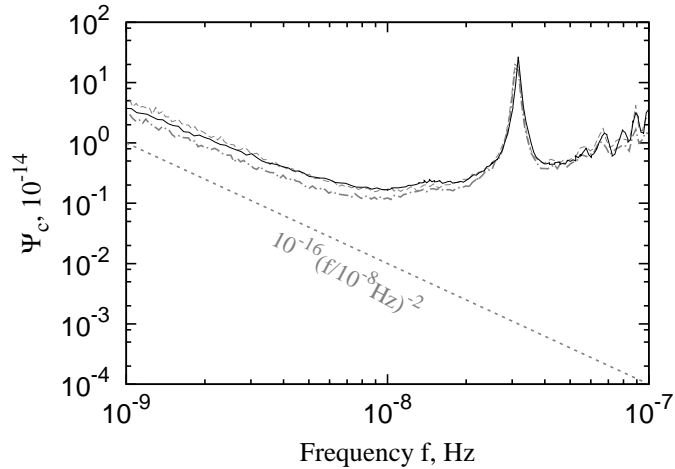}
\caption{An upper limit on the amplitude of the variable gravitational
potential \(\Psi_{c}\) due to the massive scalar field oscillations as a function of the central frequency $f$. Shown is the case of the narrow-band signal approximation (the black line),  the monochromatic signal approximation with the Earth term only (the thin gray dashed line), and
the monochromatic approximation 
using both the Earth and pulsar terms (the gray dashed-dot line); the lines are shown for the 95\% confidence level. Data from eight pulsars from the NANOGrav Project with white-noise rms residuals were used.
The dashed line shows the model amplitude (\ref{e:modelpsi}).
}
\label{result8}
\end{figure}

\begin{figure}[htp]
\includegraphics[width=.49\textwidth]{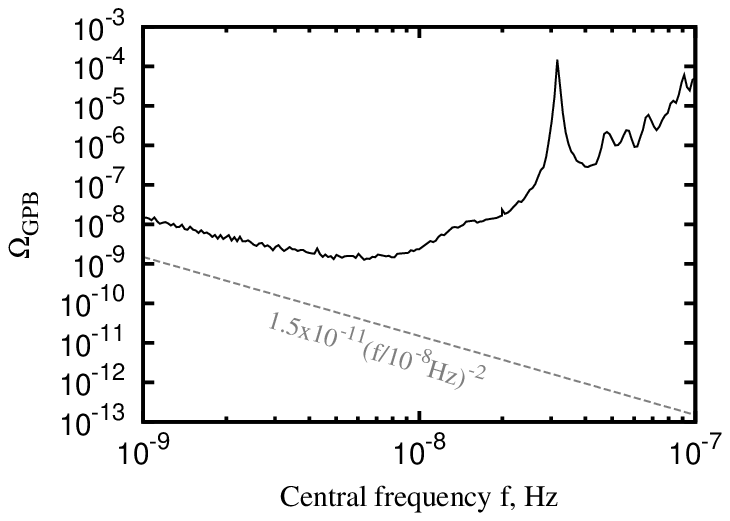}
\hfill
\includegraphics[width=.49\linewidth]{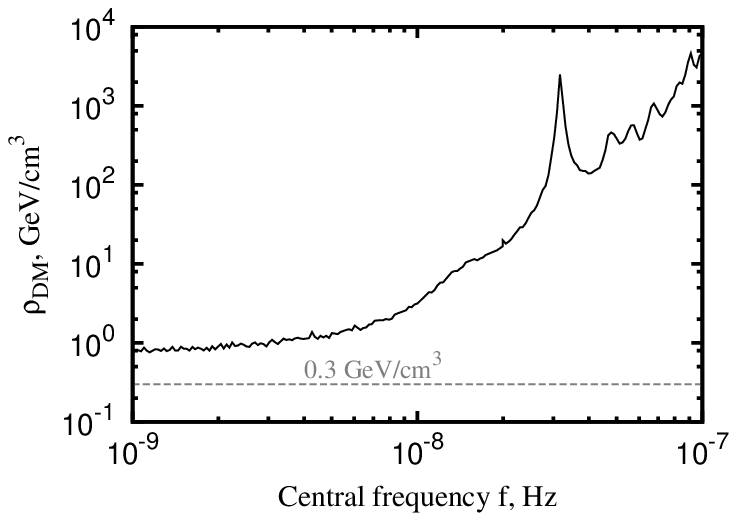}
\caption{
Left panel: An upper limit on the \(\Omega_{\mathrm{GPB}}\) 
of the ultralight scalar field as a function of frequency $f$; the solid curve 
corresponds to the 95\% confidence level. The dashed line shows the model value (\ref{e:modelomega}). 
Right panel: The same 
limit in terms of the local dark matter density \(\rho_{\text{DM}}\). 
The dashed line shows the local galactic dark matter density 0.3~GeV~cm$^{-3}$. 
Data from 12 pulsars from the NANOGrav Project have been used. 
}
\label{result12}
\end{figure}

In the monochromatic approximation (\ref{signal}), the pulsar-timing data 
are found to be more sensitive when both the 
Earth and pulsar terms are included, which is
likely to be due to the exceeding median value of the amplitude \(A_{E+P}\):
\begin{equation}
\frac{M[A_{E+P}]}{M[A_E]}=\frac{\int{2 \mathrm{cos}(\phi) \omega(\phi)}d \phi}{1}=\frac{4}{\pi}.
\end{equation}
The stringent limit obtained is \(\Psi_c<1.14 \times 10^{-15}\), 
corresponding to $h_c=2\sqrt{3}\Psi_c< 4\times 10^{-15}$ at \(f=1.75\times10^{-8}\mathrm{Hz}\) (see Fig. \(\ref{result8}\)).

 In the narrow-band approximation the power spectral density of the GBP was assumed to have a deltalike form (\ref{e:delta}). Using a flat prior in  the logarithmic scale, we numerically estimated the posterior distribution of the signal power in each frequency bin to set an upper limit on the GPB (in terms of $\Omega_{\mathrm{GPB}}$). In this case, the stringent limit is \(\Omega_{\mathrm{GPB}}<1.27 \times 10^{-9}\) at \(f=6.2\times 10^{-9}\)Hz, which corresponds
to $\Psi_c< 1.5\times 10^{-15}$ (see Fig. \(\ref{result12}\)).

\section{Conclusions}
\label{sec:discussion}

The nature of dark matter remains unclear. 
An ultralight scalar field, which can be a possible warm dark matter candidate, 
produces an oscillating pressure at a frequency $\sim m$, which via the gravitational
coupling leads to the time-variable gravitational potentials in the galactic halo. 
For electromagnetic signals propagating through time-dependent spacetime
(the galactic Sachs-Wolfe effect), these oscillations can be treated as 
a narrow-band stochastic background and thus can  
be probed in the current pulsar-timing data \cite{Khmelitsky}, opening new avenues 
for experimental tests of the possible dark matter candidates.

In the model \cite{Khmelitsky}, the dimensionless 
amplitude of the variable gravitational potential produced by
the oscillating massive scalar field \(\Psi_c\) is related to the 
local galactic dark matter density $\rho_{\text{DM}}$ and the field mass $m$ as:
\begin{equation}
\label{e:modelpsi}
\Psi_c=\pi  \frac{G\rho_{\text{DM}}}{(\pi f)^2}\approx 
10^{-16}\left(\frac{f}{10^{-8}\text{Hz}}\right)^{-2}
\approx 4.3\times 10^{-16}\left(\frac{m}{10^{-23}\text{eV}}\right)^{-2}
\left(\frac{\rho_{DM}}{0.3\text{GeV}\,\hbox{cm}^{-3}}\right)\,.
\end{equation}
In terms of the dimensionless energy density of the background (\ref{e:OmegaGPB}), we can write
\begin{equation}
\label{e:modelomega}
\Omega_{\mathrm{GPB}}\approx 1.5\times 10^{-11} \left(\cfrac{f}{10^{-8}\text{Hz}}\right)^{-2}
\left(\cfrac{\rho_{\text{DM}}}{0.3\hbox{GeV}\,\hbox{cm}^{-3}}\right)^2\,.
\end{equation}

The analysis of the NANOGrav PTA data allows us to put constraints on 
the amplitude of this signal in the monochromatic and narrow-band approximations, 
which are found to be about 1 order of magnitude higher than the predicted values (\ref{e:modelpsi}) and (\ref{e:modelomega}).
The obtained upper limits (Figs. \ref{result8} and \ref{result12}) are similar in 
both approximations due to a particularly narrow frequency range of the stochastic signal [less than one frequency bin $\Delta f\sim 1/(5\, \hbox{yr})$]. Still, the narrow-band approach 
for the analysis of pulsar-timing residuals,
as described in Sec. \ref{s:nB}
can be useful in searching for possible 
stochastic signals with a broader spectral width \(\delta f/f<1\).

Therefore, the current PTA data do not constrain the 
warm dark matter model discussed in Ref. \cite{Khmelitsky} in the phenomenologically interesting 
scalar field mass range $10^{-23}-2.3\times 10^{-23}$~eV, corresponding to 
the gravitational potential oscillation frequency range $\sim (5-12)$~nHz. Like in the case of 
monochromatic and burst GW signals, 
the sensitivity of the PTA technique to the 
specific stochastic narrow-band GBP produced by an oscillating massive scalar field should be 
determined by the rms of timing residuals of individual pulsars, unlike broadband GW backgrounds,
the sensitivity to which is mostly determined by the number of PTA pulsars \cite{2013CQGra..30v4015S}.
Thus, adding new pulsars with small rms TOA residuals into the analysis can be crucial to 
obtaining sensitive constraints on the considered model \cite{Khmelitsky} before future projects like Square Kilometre Array \cite{2010CQGra..27h4016S} become operational.

\acknowledgments{The authors thank S. Babak, M. Pshirkov, V. Rubakov and the Department of Gravitational Measurements of Sternberg Astronomical Institute for discussions and anonymous 
referees for useful notes. 
The use of the publically available NANOGrav PTA data is acknowledged.  
The work 
is supported by the Russian Science Foundation grant 14-12-00203.}

\appendix
\section{PHOTON REDSHIFT FOR SACHS-WOLFE SCALAR PERTURBATIONS}
\label{A0}

As is well known, only tensor perturbations (gravitational waves) 
cannot freely propagate in free spacetime.
To better see the difference between the effect of a gravitational wave and
a variable scalar field in the pulsar-timing, it is instructive to remind the reader
how the frequency shift appears for a photon propagating   
in spacetime in the presence of a variable massive scalar field
(the Sachs-Wolfe effect for scalar perturbations). In the covariant Newtonian gauge 
\begin{equation}
ds^2=(1+2\Phi(\bm{x},t))dt^2-(1-2\Psi(\bm{x},t))\delta_{ij}dx^idx^j
\end{equation} 
the relative frequency shift of a signal emitted at time $t'$ and received at time $t''$ 
in the linear approximation reads
(see \cite{2011iteu.book.....G} for the derivation)
\begin{equation}
\frac{\nu(t'')-\nu(t')}{\nu(t')}=\int_{t'}^{t''}(\partial_t\Phi+\partial_t\Psi)dt+\Phi(t')-\Phi(t'')\,.
\end{equation}
Here $c=1$ and the integral is taken along the unperturbed geodesic $ds^2=dt^2$.
[Note that as in the Newtonian limit, $\Phi$ plays the role of the Newtonian gravitational
potential; in a stationary spacetime, this formula expresses the standard gravitational 
redshift of a photon emitted at the point with gravitational potential $\Phi(\bm{x_{em}},t')$ and 
received at the point with gravitational potential $\Phi(\bm{x_{obs}},t'')$.] Changing from partial to full derivative in the first term, 
$\partial_t=d/dt-n_i\partial_i$, and integrating yields:
\begin{equation}
\frac{\nu(t'')-\nu(t')}{\nu(t')}=\Psi(\bm{x_{obs}},t'')-\Psi(\bm{x_{em}},t')-
\int_{t'}^{t''}n_i\partial_i(\Phi+\Psi)dt\,.
\end{equation}
This is Eq.(3.2) in Ref. \cite{Khmelitsky}. 
To see that the second term is small, one can take, for example, expression
with changing phase and amplitude and integrate along the trajectory $x=t$:
\begin{equation}
\int_{t'}^{t''}\partial_x\left(kx(t)\sin(\omega t+kx(t))\right)dt=
\left.\left(-\frac{k\omega}{(\omega+k)^2}\cos(\omega k+t)+ 
\frac{k^2t}{\omega+k}\sin(\omega k+t) \right) \right|_{t'}^{t''}\,.
\end{equation}
Even if $k(t''-t')\sim 1$ and the integrand strongly changes along the
trajectory, the result is suppressed by the small factor $k/\omega=v\sim 10^{-3}$ 
relative to the value of the potentials. It can be easily taken into account in the
PTA data analysis, its contribution being attributed to the field phase uncertainty.

\section{PULSAR INTRINSIC NOISE}
\label{A}
\begin{figure*}[htp]
\centering
\includegraphics[scale=0.50]{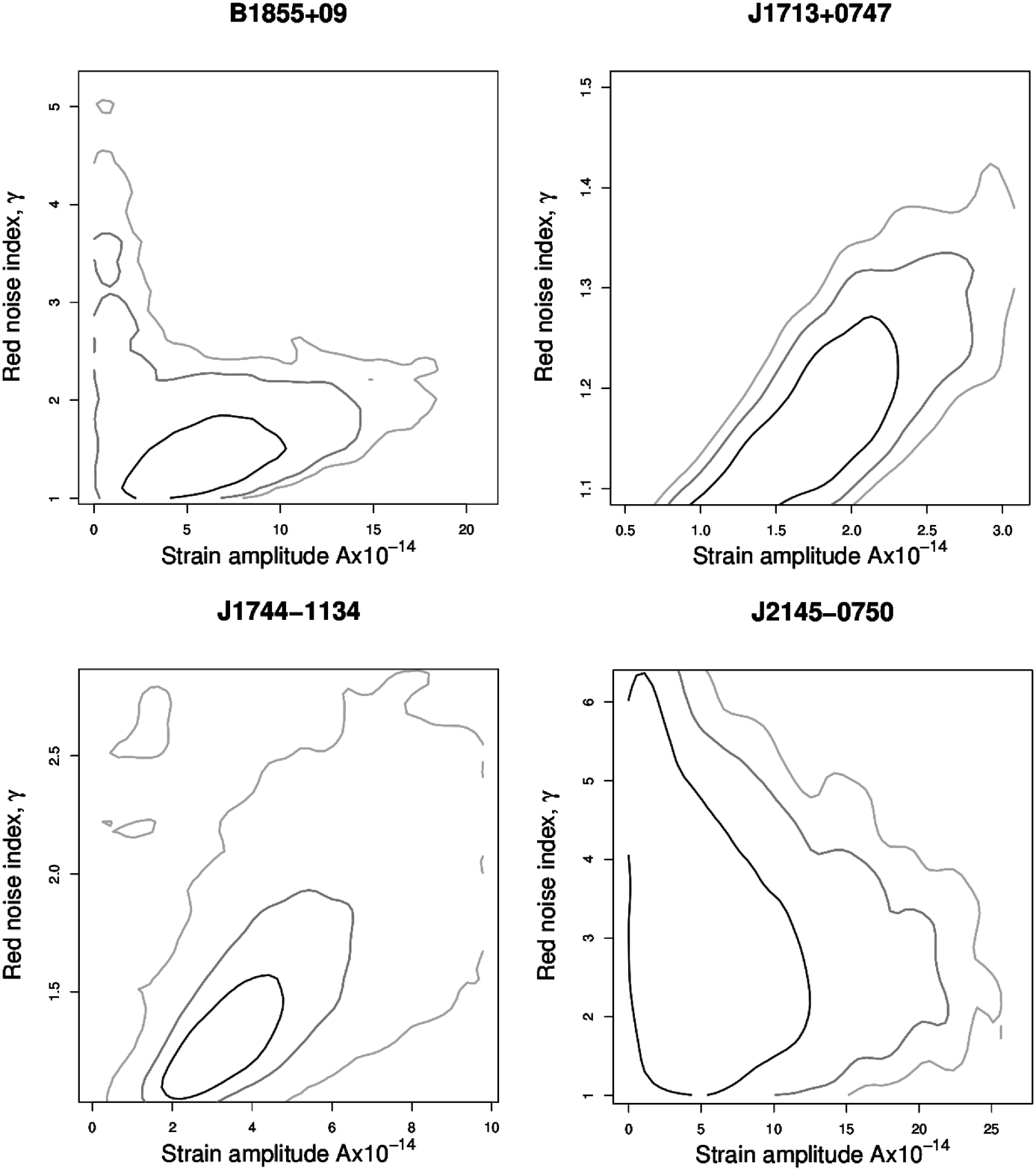}
\caption{Estimation of the red-noise parameters \(A_{RN}\) and \(\gamma_{RN}\) 
for pulsars J1713+0747, J2145-0750, B1855+09 and J1744-1134 using the Markov chain Monte-Carlo method. Different confidence levels are shown in color: black, 68 \% confidence level; gray, 95\% confidence level; and light gary, 99.7 \% confidence level. The plots were obtained using the $R$ statistical package \footnote{\url{http://www.r-project.org/}}.}
\label{f:rednoise}
\end{figure*}

The intrinsic pulsar red noise is a challenging problem in the pulsar-timing analysis
because it strongly affects the PTA sensitivity
to GW signals. The nature of this type of noise is not completely clear and can be 
related, for example, to irregular momentum exchange between the superfluid component and the crust 
of the neutron star, as well as with fluctuations of the electron density in the interstellar medium \cite{Lommen}. Therefore, the red noise should 
definitely be included in the signal model in the data analysis.

The red-noise spectrum is usually assumed to have a power-law form
\begin{equation}
S(f)=\frac{A^2}{12 \pi^2} f_0^3 \left(\frac{f}{f_{0}}\right)^{-\gamma}.
\end{equation}
In the time domain, the Wiener-Khinchin theorem allows us to obtain the covariance function presented by Eq. (\ref{C}).
The red-noise component characterized by two parameters \(A_{RN}\) and \(\gamma_{RN}\) was estimated individually for each of four pulsars (J1713+0747, J2145-0750, B1855+09 and J1744-1134) by the numerical estimation of the probability distribution from MCMC simulations. The results are
presented in Fig. \ref{f:rednoise}.
The estimated red-noise parameters for these pulsars have been 
included in further data analysis 
to obtain the final results shown in Fig. \(\ref{result12}\).

\section{COVARIANCE MATRIX FOR GPB}
\label{B}
The covariance matrix of a stochastic process can be derived from its power spectral density using the Wiener-Khinchin theorem:
\begin{equation}
C(\tau)=\int^{\infty}_{0}S(f)\cos (\tau f)df.
\end{equation}
In our case, the \(S(f)\) has the form [see Eq. \(\ref{Spectr}\)]: 
\begin{equation}
S(f)=\frac{Q}{f^3},
\end{equation}
where $Q$ is some constant; therefore, the following procedure can be applied. 
Let us expand \(\cos(\tau f)\) in a Maclaurin series:
\begin{equation}
\cos(\tau f)=\sum_{n=0}^{\inf}\frac{(-1)^n}{(2n)!}(\tau f)^{2n}, x \in \mathbb{C}.
\end{equation}
After performing the integral, we get:
\begin{equation}
C(\tau)=\sum_{n=0}^{\inf}\frac{(-1)^n}{(2n)!(2n-2)}(\tau)^{2n}f^{2n-2}((1+\frac{\delta f}{f})^{2n-2}-1).
\end{equation}
In the narrow-band approximation $\delta f/f\ll 1$, by expanding \((1+\frac{\delta f}{f})^{2n-2}\) in a Maclaurin series we find:
\begin{equation}
C(\tau)=\frac{Q}{f^2}\left\{\cos(f\tau)\left(\frac{\delta f}{f}\right)+(-3\cos(f\tau)-f\tau\sin(f \tau))\left(\frac{\delta f}{f}\right)^2+O\left(\left(\frac{\delta f}{f}\right)^3\right)\right\}.
\end{equation}
A very narrow frequency range of the sought signal allows us to retain  
only the first-order terms.

\bibliography{gw.bib}
\end{document}